\documentstyle[aps,eqsecnum,preprint,floats,epsf,epsfig]{revtex}
\begin{document}
\def\be{\begin{eqnarray}}
\def\en{\end{eqnarray}}
\def\non{\nonumber}
\def\la{\langle}
\def\ra{\rangle}
\def\nc{N_c^{\rm eff}}
\def\vp{\varepsilon}
\def\vma{{_{V-A}}}
\def\vpa{{_{V+A}}}
\def\m{\hat{m}}
\def\ov{\overline}
\def\pr{{\sl Phys. Rev.}~}
\def\prl{{\sl Phys. Rev. Lett.}~}
\def\pl{{\sl Phys. Lett.}~}
\def\np{{\sl Nucl. Phys.}~}
\def\zp{{\sl Z. Phys.}~}
\def\lsim{ {\ \lower-1.2pt\vbox{\hbox{\rlap{$<$}\lower5pt\vbox{\hbox{$\sim$}
}}}\ } }
\def\gsim{ {\ \lower-1.2pt\vbox{\hbox{\rlap{$>$}\lower5pt\vbox{\hbox{$\sim$}
}}}\ } }

\font\el=cmbx10 scaled \magstep2 {\obeylines \hfill October, 1999}

\vskip 1.5 cm

\begin{center}
{\large {\bf Factorization Approach for the $\Delta I=1/2$
Rule}}\par {\large {\bf and $\varepsilon'/\varepsilon$ in Kaon
Decays }}

\vskip 1.0cm {\bf Hai-Yang Cheng} \vskip 0.5cm Institute of
Physics, Academia Sinica,\par Taipei, Taiwan 115, Republic of
China
\end{center}
\vskip 1.0cm

\centerline{\bf Abstract} \vskip 0.3cm {\small The $\Delta I=1/2$
rule and direct CP violation $\varepsilon'/\varepsilon$ in kaon
decays are studied within the framework of the effective
Hamiltonian approach in conjunction with generalized factorization
for hadronic matrix elements. We identify two principal sources
responsible for the enhancement of ${\rm Re}A_0/{\rm Re}A_2$: the
vertex-type as well as penguin-type corrections to the matrix
elements of four-quark operators, which render the physical
amplitude renormalization-scale and -scheme independent, and the
nonfactorized effect due to soft-gluon exchange, which is needed
to suppress the $\Delta I=3/2$ $K\to\pi\pi$ amplitude. Contrary to
the chiral approach which is limited to light meson decays and
fails to reproduce the $A_2$ amplitude, the aforementioned
approach for dealing with scheme and scale issues is applicable to
heavy meson decays. We obtain ${\rm Re}A_0/{\rm Re}A_2=13-15$ if
$m_s$(1\,GeV) lies in the range (125$-$175)\,MeV. The bag
parameters $B_i$, which are often employed to parametrize the
scale and scheme dependence of hadronic matrix elements, are
calculated in two different renormalization scehemes. It is found
that $B_8^{(2)}$ and $ B_6^{(0)}$, both of order 1.5 at $\mu=1$
GeV, are nearly $\gamma_5$ scheme independent, whereas
$B^{(0)}_{3,5,7}$ as well as $B_7^{(2)}$ show a sizable scheme
dependence. Moreover, only $B_{1,3,4}^{(0)}$ exhibit a significant
$m_s$ dependence, while the other $B$-parameters are almost $m_s$
independent. For direct CP violation, we obtain
$\varepsilon'/\varepsilon=(0.7-1.1)\times 10^{-3}$ if $m_s(1\,{\rm
GeV})=150$ MeV and $\varepsilon'/\varepsilon=(1.0-1.6)\times
10^{-3}$ if $m_s$ is as small as indicated by some recent lattice
calculations.

}

\pagebreak

\section{Introduction}
The celebrated $\Delta I=1/2$ rule in kaon decays still remains an
enigma after the first observation more than four decades ago. The
tantalizing puzzle is the problem of how to enhance the $A_0/A_2$
ratio of the $\Delta I=1/2$ to $\Delta I=3/2$ $K\to\pi\pi$
amplitudes from the outrageously small value 0.9 [see Eq.
(\ref{naive}) below] to the observed value $22.2\pm 0.1$ (for a
review of the $\Delta I=1/2$ rule, see \cite{Cheng89}). Within the
framework of the effective weak Hamiltonian in conjunction with
the factorization approach for hadronic matrix elements, the
$A_0/A_2$ ratio is at most of order 8 even after the nonfactorized
soft-gluon effects are included \cite{Cheng89}. Moreover, the
$\mu$ dependence of hadronic matrix elements is not addressed in
the conventional calculation. In the past ten years or so, most
efforts are devoted to computing the matrix elements to ${\cal
O}(p^4)$ in chiral expansion. This scenario has the advantages
that chiral loops introduces a scale dependence for hadronic
matrix elements and that meson loop contributions to the $A_0$
amplitude are large enough to accommodate the data. However, this
approach also exists a fundamental problem, namely the
long-distance evolution of meson loop contributions can only be
extended to the scale of order 600 MeV, whereas the perturbative
evaluation of Wilson coefficients cannot be reliably evolved down
to the scale below 1 GeV. The conventional practice of matching
chiral loop corrections to hadronic matrix elements with Wilson
coefficient functions at the scale $\mu=(0.6-1.0)$ GeV requires
chiral perturbation theory and/or perturbative QCD be pushed into
the regions beyond their applicability.

Another serious difficulty with the chiral approach is that
although the inclusion of chiral loops will make a large
enhancement for $A_0$, it cannot explain the $A_2$ amplitude. For
example, in the analysis of \cite{Hambye} in which a physical
cutoff $\Lambda_c$ is introduced to regularize the quadratic and
logarithmic divergence of the long-distance chiral loop
corrections to $K\to\pi\pi$ amplitudes, the amplitude $A_2$ is
predicted to be highly unstable relative to the cutoff scale
$\Lambda_c$ and it even changes sign at $\Lambda_c\gsim 650$ MeV
\cite{Hambye,Bijnens}. In the approach in which the dimensional
regularization is applied to regularize the chiral loop
divergences and to consistently match the logarithmic scale
dependence of Wilson coefficients, the predicted $A_2$ amplitude
is too large compared to experiment \cite{Kambor}, indicating the
necessity of incorporating nonfactorized effects to suppress the
$\Delta I=3/2$ amplitude \cite{Antonelli}. This implies that not
all the long-distance nonfactorized contributions to hadronic
matrix elements are fully accounted for by chiral loops.  In
short, it is not possible to reproduce $A_0$ and $A_2$ amplitudes
{\it simultaneously} by chiral loops alone.

Even if the scale dependence of $K\to\pi\pi$ matrix elements can
be furnished by meson loops, it is clear that this approach based
on chiral perturbation theory is not applicable to heavy meson
decays. Therefore, it is strongly desirable to describe the
nonleptonic decays of kaons and heavy mesons within the same
framework.

In the effective Hamiltonian approach, the renormalization scale
and scheme dependence of Wilson coefficients is compensated by
that of the hadronic matrix elements of four-quark operators
$O(\mu)$ renormalized at the scale $\mu$. Since there is no
first-principles evaluation of $\la O(\mu)\ra$ except for lattice
calculations, it becomes necessary to compute the vertex- and
penguin-type corrections to $\la O\ra$ (not $\la O(\mu)\ra$ !),
which account for the scale and scheme dependence of $\la
O(\mu)\ra$, and then apply other methods to calculate $\la O\ra$.
The $\Delta I=1/2$ rule arises from the cumulative effects of the
short-distance Wilson coefficients, penguin operators, final-state
interactions, nonfactorized effects due to soft-gluon exchange,
and radiative corrections to the matrix elements of four-quark
operators. As shown in \cite{Cheng99}, the last two effects are
the main ingredients for the large enhancement of $A_0$ with
respect to $A_2$.

Contrary to the nonfactorized effects in charmless $B$ decays,
which are dominated by hard gluon exchange in the heavy quark
limit \cite{Beneke} and expected to be small due to the large
energy released in the decay process, the nonfactorized term in
$K\to\pi\pi$ is anticipated to be large and nonperturbative in
nature, namely it comes mainly from soft gluon exchange. One can
use $K^+\to\pi^+\pi^0$ to extract the nonfactorizable
contributions to the hadronic matrix elements of $(V-A)(V-A)$
four-quark operators \cite{Cheng99}.

Instead of using scheme- and scale-independent effective Wilson
coefficients, one can alternatively parametrize the hadronic
matrix elements in terms of the bag parameters $B_i^{(0)}$ and
$B_i^{(2)}$ which describe the scale and scheme dependence of
hadronic matrix elements $\la Q_i(\mu)\ra$. These non-perturbative
parameters are evaluated in the present paper. We have checked
explicitly that these two seemingly different approaches yield the
same results.

The prediction of $\varepsilon'/\varepsilon$ in the standard model
is often plagued by the difficulty that the result depends on the
choice of the renormalization scheme. Presumably this is not an
issue in the effective Wilson coefficient approach. Unfortunately,
as we shall see in Sec. IV, our predictions for
$\varepsilon'/\varepsilon$ are scheme dependent for reasons not
clear to us.

The present paper is organized as follows. In Sec. II we construct
scheme and scale independent effective Wilson coefficients
relevant to $K\to\pi\pi$ decays and direct CP violation
$\varepsilon'/\varepsilon$. The bag parameters $B_i$ are evaluated
in Sec. III. Based on the effective Wilson coefficients or bag
parameters, $K\to\pi\pi$ amplitudes and direct CP violation are
calculated in Sec. IV and their results are discussed in Sec. V.
Sec. VI is for the conclusion.

\section{effective Wilson coefficients}
The effective Hamiltonian relevant to $K\to\pi\pi$ transition is
\be
{\cal H}_{\rm eff}(\Delta S=1) &=&
{G_F\over\sqrt{2}}V_{ud}V_{us}^*
\Bigg(\sum^{10}_{i=1}c_i(\mu)Q_i(\mu)\Bigg)+{\rm h.c.},
\en
where
\be
c_i(\mu)=z_i(\mu)+\tau y_i(\mu),
\en
with $\tau=-V_{td}V_{ts}^*/(V_{ud}V_{us}^*)$, and \be  \label{Oi}
&& Q_1= (\bar ud)_\vma(\bar su)_\vma, \qquad\qquad\qquad\qquad~
Q_2 = (\bar u_\alpha b_\beta)_\vma(\bar q_\beta u_\alpha)_\vma,
\non \\
 && Q_{3(5)}=(\bar sd)_\vma\sum_{q}(\bar qq)_{\vma(\vpa)}, \qquad
\qquad Q_{4(6)}=(\bar s_\alpha d_\beta)_\vma\sum_{q}(\bar q_\beta
q_\alpha)_{ \vma(\vpa)},   \\ && Q_{7(9)}={3\over 2}(\bar
sd)_\vma\sum_{q}e_{q}(\bar qq)_{\vpa(\vma)},
  \qquad~ Q_{8(10)}={3\over 2}(\bar s_\alpha d_\beta)_\vma\sum_{q}
  e_{q}(\bar q_\beta q_\alpha)_{\vpa(\vma)},  \non
\en
with $Q_3$--$Q_6$ being the QCD penguin operators,
$Q_{7}$--$Q_{10}$ the electroweak penguin operators and $(\bar
q_1q_2)_{_{V\pm A}}\equiv\bar q_1\gamma_\mu(1\pm \gamma_5)q_2$.
The sum in Eq. (\ref{Oi}) is over light flavors, $q=u,d,s$.

In the absence of first-principles calculations for hadronic
matrix elements, it is customary to evaluate the matrix elements
under the factorization hypothesis so that $\la Q(\mu)\ra$ is
factorized into the product of two matrix elements of single
currents, governed by decay constants and form factors. However,
the naive factorized amplitude is not renormalization scale- and
$\gamma_5$ scheme- independent as the scale and scheme dependences
of Wilson coefficients are not compensated by that of the
factorized hadronic matrix elements. In principle, the scale and
scheme problems with naive factorization will not occur in the
full amplitude since $\la Q(\mu)\ra$ involves vertex-type and
penguin-type corrections to the hadronic matrix elements of the
4-quark operator renormalized at the scale $\mu$. Schematically,
\be
{\rm weak~decay~amplitude} &=& {\rm
naive~factorization~+~vertex\!-\!type~corrections} \\ &+&{\rm
penguin\!-\!type~corrections~+~spectator~contributions}+\cdots,
\non
\en
where the spectator contributions take into account the gluonic
interactions between the spectator quark of the kaon and the
outgoing light meson. The perturbative part of vertex-type and
penguin-type corrections will render the decay amplitude scale and
scheme independent. Generally speaking, the Wilson coefficient
$c(\mu)$ takes into account the physics evolved from the scale
$M_W$ down to $\mu$, while $\la Q(\mu)\ra$ involves evolution from
$\mu$ down to the infrared scale. Formally, one can write
\be
\la Q(\mu)\ra=g(\mu,\mu_f)\la Q(\mu_f)\ra,
\en
where $\mu_f$ is a factorization scale, and $g(\mu,\mu_f)$ is an
evolution factor running from the scale $\mu$ to $\mu_f$ which is
calculable because the infrared structure of the amplitude is
absorbed into $\la Q(\mu_f)\ra$. Writing
\be
c^{\rm eff}(\mu_f)=c(\mu)g(\mu,\mu_f),
\en
the effective Wilson coefficients will be scheme and $\mu$-scale
independent. Of course, it appears that the $\mu$-scale problem
with naive factorization is traded in by the $\mu_f$-scale
problem. Nevertheless, once the factorization scale at which we
apply the factorization approximation to matrix elements is fixed,
the physical amplitude is independent of the choice of $\mu$. More
importantly, the effective Wilson coefficients are
$\gamma_5$-scheme independent. In principle, one can work with any
quark configuration, on-shell or off-shell, to compute the full
amplitude. Note that if external quarks are off-shell and if the
off-shell quark momentum is chosen as the infrared cutoff,
$g(\mu,\mu_f)$ will depend on the gauge of the gluon field
\cite{Buras98}. But this is not a problem at all since the gauge
dependence belongs to the infrared structure of the wave function.
However, if factorization is applied to $\la Q(\mu_f)\ra$, the
information of the gauge dependence characterized by the wave
function will be lost. Hence, as stressed in \cite{CLY,CCTY}, in
order to apply factorization to matrix elements and in the
meantime avoid the gauge problem connected with effective Wilson
coefficients, one must work in the on-shell scheme to obtain gauge
invariant and infrared finite $c_i^{\rm eff}$ and then applies
factorization to $\la Q(\mu_f)\ra$ afterwards. Of course, physics
should be $\mu_f$ independent. In the formalism of the
perturbative QCD factorization theorem, the nonperturbative meson
wave functions are specified with the dependence on the
factorization scale $\mu_f$ \cite{CLY}. These wave functions are
universal for all decay processes involving the same mesons.
Hence, a consistent evaluation of hadronic matrix elements will
eventually resort to the above-mentioned meson wave functions
determined at the scale $\mu_f$.

In general, the scheme- and $\mu$-scale-independent effective
Wilson coefficients have the form \cite{Ali,CT98}:
\be
\label{ceff1} c_i^{\rm eff}(\mu_f) &=& c_i(\mu)+{\alpha_s\over
4\pi}\left(\gamma_V^{T}\ln{\mu_f\over \mu}+\hat
r_V^T\right)_{ij}c_j(\mu)+~{\rm penguin\!-\!type~corrections},
\en
where $\mu_f$ is the factorization scale arising from the
dimensional regularization of infrared divergence \cite{CLY}, and
the anomalous dimension matrix $\gamma_V$ as well as the constant
matrix $\hat r_V$ arise from the vertex-type corrections to
four-quark operators. For kaon decays under consideration, there
is no any heavy quark mass scale between $m_c$ and $m_K$. Hence,
the logarithmic term emerged in the vertex corrections to 4-quark
operators is of the form $\ln(\mu_f/\mu)$ as shown in Eq.
(\ref{ceff1}). We will set $\mu_f=1$ GeV in order to have a
reliable estimate of perturbative effects on effective Wilson
coefficients.

It is known that the penguin operators $Q_{5,6}$ do not induce
$K^0\to\pi\pi$ directly, but their Fierz transformations via
$(V-A)(V+A)\to -2(S+P)(S-P)$ do make contributions. Applying
equations of motion, $\la Q_{5,6}(\mu)\ra$ are proportional to
$m_K^2/[m_s(\mu)+m_q(\mu)]$ with $q=u$ or $d$. This means that,
contrary to current$\times$current operators, the matrix elements
$\la Q_{5,6}(\mu)\ra$ for $K-\pi\pi$ transition under the vacuum
insertion approximation {\it do} exhibit a $\mu$ dependence
governed by light quark masses. The $\mu$ dependence of the Wilson
coefficients $c_{5,6}(\mu)$ is essentially compensated by that of
light quark masses (the cancellation becomes exact in the
large-$N_c$ limit). Of course, the near cancellation of $\mu$
dependence does not imply that factorization works for the matrix
elements of density$\times$density operators since the scheme
dependence of $c_{5,6}(\mu)$ still does not get compensation. It
is thus advantageous to apply the aforementioned effective Wilson
coefficients to avoid the scheme problem caused by factorization.
And in the meantime, the $\mu_f$ dependence of $c^{\rm
eff}_{5,6}(\mu_f)$ is largely canceled by that of quark masses
entering the matrix elements $\la Q_{5,6}(\mu_f)\ra$.

To proceed, we note that the renormalization-scale and -scheme
independent effective Wilson coefficient functions $\tilde
z_i^{\rm eff}$ are given by (for details, see \cite{CLY,CCTY}):
\be
\tilde z_1^{\rm eff} &=& z_1(\mu)+{\alpha_s\over
4\pi}\left(\gamma_V^{T}\ln{\mu_f\over \mu}+\hat
r_V^T\right)_{1i}z_i(\mu), \non \\
\tilde z_2^{\rm eff} &=&
z_2(\mu)+{\alpha_s\over 4\pi}\left(\gamma_V^{T}\ln{\mu_f\over
\mu}+\hat r_V^T\right)_{2i}z_i(\mu), \non \\
\tilde z_3^{\rm eff}
&=& z_3(\mu)+{\alpha_s\over 4\pi}\left(\gamma_V^{T}
\ln{\mu_f\over \mu}+\hat r_V^T\right)_{3i}z_i(\mu)-{\alpha_s \over
24\pi}(C_t+C_p), \non \\
\tilde z_4^{\rm eff} &=&
z_4(\mu)+{\alpha_s\over 4\pi}\left(\gamma_V^{T}\ln{\mu_f\over
\mu}+\hat r_V^T\right)_{4i}z_i(\mu)+{\alpha_s\over 8\pi}(C_t+C_p),
\non \\
\tilde z_5^{\rm eff} &=& z_5(\mu)+{\alpha_s\over
4\pi}\left(\gamma_V^{T}\ln{\mu_f\over \mu}+\hat
r_V^T\right)_{5i}z_i(\mu)-{\alpha_s\over 24\pi}(C_t+C_p), \non \\
\tilde z_6^{\rm eff} &=& z_6(\mu)+{\alpha_s\over
4\pi}\left(\gamma_V^{T}\ln{\mu_f\over \mu}+\hat
r_V^T\right)_{6i}z_i(\mu)+{\alpha_s\over 8\pi}(C_t+C_p), \non \\
\tilde z_7^{\rm eff} &=& z_7(\mu)+{\alpha_s\over
4\pi}\left(\gamma_V^{T}\ln{\mu_f\over \mu}+\hat
r_V^T\right)_{7i}z_i(\mu)+{\alpha\over 8\pi}C_e, \non \\ \tilde
z_8^{\rm eff} &=& z_8(\mu)+{\alpha_s\over
4\pi}\left(\gamma_V^{T}\ln{\mu_f\over \mu}+\hat
r_V^T\right)_{8i}z_i(\mu), \non \\
\tilde z_9^{\rm eff} &=&
z_9(\mu)+{\alpha_s\over 4\pi}\left(\gamma_V^{T} \ln{\mu_f\over
\mu}+\hat r_V^T\right)_{9i}z_i(\mu)+{\alpha\over 8\pi}C_e, \non \\
\tilde z_{10}^{\rm eff} &=& z_{10}(\mu)+{\alpha_s\over
4\pi}\left(\gamma_V^{T}\ln{\mu_f\over \mu}+\hat
r_V^T\right)_{10i}z_i(\mu). \label{zeff}
\en
Likewise, the effective Wilson coefficients $\tilde y_i^{\rm eff}$
have similar expressions in terms of $y_i(\mu)$ except that
$y_1=y_2=0$. The reason why we call the l.h.s. of Eq. (\ref{zeff})
as  $\tilde z_i^{\rm eff}$ rather than $z_i^{\rm eff}$ will become
shortly. In Eq. (\ref{zeff}) the superscript $T$ denotes a
transpose of the matrix, the anomalous dimension matrix
$\gamma_V$ as well as the constant matrix $\hat r_V$ arise
from the vertex corrections to the operators $Q_1-Q_{10}$, $C_t$,
$C_p$ and $C_e$ from the QCD penguin-type diagrams of the
operators $Q_{1,2}$, the QCD penguin-type diagrams of the
operators $Q_3-Q_6$, and the electroweak penguin-type diagram of
$Q_{1,2}$, respectively:
\be
C_t &=& \left({2\over 3}\kappa-G(m_u,k,\mu)\right)z_1, \non \\ C_p
&=& \left( {4\over
3}\kappa-G(m_s,k,\mu)-G(m_d,k,\mu)\right)z_3-\sum_{i=u,d,s}
G(m_i)(z_4+z_6), \non \\ C_e &=& {8\over 9}\left({2\over
3}\kappa-G(m_u,k,\mu)\right)(z_1+3z_2),
\en
where $\kappa$ is a parameter characterizing the $\gamma_5$-scheme
dependence in dimensional regularization, i.e.,
\be
\kappa=\cases{ 1 & NDR,  \cr 0 & HV,  \cr}
\en
in the naive dimensional regularization (NDR) and 't Hooft-Veltman
(HV) schemes for $\gamma_5$, and the function $G(m,k,\mu)$ is
given by
\be
\label{G} G(m,k,\mu)=-4\int^1_0dx\,x(1-x)\ln\left(
{m^2-k^2x(1-x)\over \mu^2}\right),
\en
with $k^2$ being the momentum squared carried by the virtual
gluon. The explicit expression for $\gamma_V$ is given in
\cite{Ali}.  For reader's convenience, we list here the constant
matrix $\hat r_V$ \cite{CCTY,Cheng99}:
\be
\label{rndr} \hat r_V^{\rm NDR}=\left(\matrix{ 3 & -9 & 0 & 0 & 0
& 0 & 0 & 0 & 0 & 0 \cr -9 & 3 & 0 & 0 & 0 & 0 & 0 & 0 & 0 & 0 \cr
0 & 0 & 3 & -9 & 0 & 0 & 0 & 0 & 0 & 0 \cr  0 & 0 & -9 & 3 & 0 & 0
& 0 & 0 & 0 & 0 \cr 0 & 0 & 0 & 0 & -1 & 3 & 0 & 0 & 0 & 0 \cr 0 &
0 & 0 & 0 & -3 & 17 & 0 & 0 & 0 & 0 \cr 0 & 0 & 0 & 0 & 0 & 0 & -1
& 3 & 0 & 0 \cr 0 & 0 & 0 & 0 & 0 & 0 & -3 & 17 & 0 & 0 \cr 0 & 0
& 0 & 0 & 0 & 0 & 0 & 0 & 3 & -9 \cr 0 & 0 & 0 & 0 & 0 & 0 & 0 & 0
& -9 & 3 \cr}\right)
\en
in the NDR scheme, and
\be
\label{rhv} \hat r_V^{\rm HV}=\left(\matrix{ {7\over 3} & -7 & 0 &
0 & 0 & 0 & 0 & 0 & 0 & 0 \cr -7 & {7\over 3} & 0 & 0 & 0 & 0 & 0
& 0 & 0 & 0 \cr 0 & 0 & {7\over 3} & -7 & 0 & 0 & 0 & 0 & 0 & 0
\cr 0 & 0 & -7 & {7\over 3} & 0 & 0 & 0 & 0 & 0 & 0 \cr 0 & 0 & 0
& 0 & -3 & 9 & 0 & 0 & 0 & 0 \cr 0 & 0 & 0 & 0 & 1 & {47\over 3} &
0 & 0 & 0 & 0 \cr 0 & 0 & 0 & 0 & 0 & 0 & -3 & 9 & 0 & 0 \cr 0 & 0
& 0 & 0 & 0 & 0 & 1 & {47\over 3} & 0 & 0 \cr 0 & 0 & 0 & 0 & 0 &
0 & 0 & 0 & {7\over 3} & -7 \cr 0 & 0 & 0 & 0 & 0 & 0 & 0 & 0 & -7
& {7\over 3} \cr}\right)
\en
in the HV scheme. Note that the 66 and 88 entries of $\hat r_V$
given in \cite{CCTY} are erroneous and have been corrected in
\cite{Cheng99} and \cite{CY99}.

The results of a direct calculation of $\tilde z_i^{\rm eff}$ and
$\tilde y_i^{\rm eff}$ in NDR and HV schemes using Eq.
(\ref{zeff}) are displayed in Table I. Formally, the effective
Wilson coefficients are scale and scheme independent up to the
order $\alpha_s$. This implies that the scheme independence of
$z_i^{\rm eff}$ requires that the Wilson coefficients $z_i(\mu)$
appearing in the vertex-type corrections and in $C_t,~C_p$ and
$C_e$ be replaced by the lowest-order (LO) ones $z_i^{\rm LO}$,
while $z_1$ appearing in $C_t$ be the next-to-leading order (NLO)
one, and likewise for $y_i^{\rm eff}$. Therefore, we define the
effective Wilson coefficients ${z}_i^{\rm eff}$ and ${y}_i^{\rm
eff}$ similar to Eq. (\ref{zeff}) except for the above-mentioned
replacement, for example,
\be
z_1^{\rm eff} &=& z_1(\mu)+{\alpha_s\over
4\pi}\left(\gamma_V^{T}\ln{\mu_f\over \mu}+\hat
r_V^T\right)_{1i}z_i^{\rm LO}(\mu).
\en
From Table I we see that the scale independence of ${z}_i^{\rm
eff}$, which is good to the accuracy of the third digit, is
significantly better than $\tilde z_i^{\rm eff}$; for example, the
values of $z_{4,6}$(NDR) and $z_{4,6}$(HV) to NLO are quite
different, but their effective Wilson coefficients $z_{4,6}^{\rm
eff}$ are obviously scheme independent. Note that $z_3^{\rm
eff},\cdots,z_6^{\rm eff}$ are enhanced relative to their NLO
values by about three times as they receive large corrections
proportional to $\alpha_sC_t$. By contrast, the scheme
independence of $y_i^{\rm eff}$ is not as good as $z_i^{\rm eff}$.
In particular, $y_6^{\rm eff}$, which plays an important role in
$\varepsilon'/\varepsilon$, shows a slight scheme dependence for
reasons not clear to us.

\vskip 0.4cm
\begin{table}[ht]
\caption{ $\Delta S=1$ Wilson coefficients at $\mu=1$ GeV for
$m_t=170$ GeV and $\Lambda_{\ov{\rm MS}}^{(4)}=325$ MeV, taken
from Table XVIII of [17]. Also shown are the effective Wilson
coefficients $\tilde z_i^{\rm eff}$ and $z_i^{\rm eff}$ (see the
text), $\tilde y_i^{\rm eff}$ and $y_i^{\rm eff}$ in NDR and HV
schemes with $\mu=1$ GeV, $\mu_f=1$ GeV and $k^2=m_K^2/2$. Note
that $y_1=y_2=0$.}
\begin{center}
\begin{tabular}{ l r r r r r r r }
 & LO & NDR & HV & $\tilde z_i^{\rm eff}$(NDR) & $\tilde z_i^{\rm eff}$(HV) &
$z_i^{\rm eff}$(NDR) & $z_i^{\rm eff}$(HV) \\ \hline
$z_1$ & 1.433 & 1.278 & 1.371 & 1.614 & 1.678 & 1.718 & 1.713 \\
$z_2$ & -0.748 & -0.509 & -0.640 & -1.029 & -1.082 & -1.113 & -1.110 \\
$z_3$ & 0.004 & 0.013 & 0.007 & 0.039 & 0.034 & 0.032 & 0.032 \\
$z_4$ & -0.012 & -0.035 & -0.017 & -0.080 & -0.085 & -0.081 & -0.084 \\
$z_5$ & 0.004 & 0.008 & 0.004 & 0.024 & 0.024 & 0.024 & 0.025 \\
$z_6$ & -0.013 & -0.035 & -0.014 & -0.094 & -0.086 & -0.086 & -0.086 \\
$z_7/\alpha$ & 0.008 & 0.011 & -0.002 & 0.025 & 0.047 & 0.063 & 0.069 \\
$z_8/\alpha$ & 0.001 & 0.014 & 0.010 & 0.025 & 0.016 & 0.016 & 0.013 \\
$z_9/\alpha$ & 0.008 & 0.018 & 0.005 & 0.039 & 0.057 & 0.072 & 0.078 \\
$z_{10}/\alpha$ & -0.001 & -0.008 & -0.010 & -0.015 & -0.012 & -0.011 & -0.012
 \\ \hline
\vspace{0.1cm}
& LO & NDR & HV & $\tilde y_i^{\rm eff}$(NDR) & $\tilde y_i^{\rm eff}$(HV) &
$y_i^{\rm eff}$(NDR) & $y_i^{\rm eff}$(HV)  \\
\hline
$y_3$ & 0.038 & 0.032 & 0.037 & 0.048 & 0.050 & 0.050 & 0.049 \\
$y_4$ & -0.061 & -0.058 & -0.061 & -0.051 & -0.055 & -0.053 & -0.053 \\
$y_5$ & 0.013 & -0.001 & 0.016 & 0.004 & 0.003 & 0.003 & 0.002 \\
$y_6$ & -0.113 & -0.111 & -0.097 & -0.161 & -0.130 & -0.160 & -0.138 \\
$y_7/\alpha$ & 0.036 & -0.032 & -0.030 & -0.051 & -0.019 & -0.052 & -0.028 \\
$y_8/\alpha$ & 0.158 & 0.173 & 0.188 & 0.287 & 0.295 & 0.285 & 0.300 \\
$y_9/\alpha$ & -1.585 & -1.576 & -1.577 & -2.013 & -1.919 & -2.053 & -1.948 \\
$y_{10}/\alpha$ & 0.800 & 0.690 & 0.699 & 1.339 & 1.205 & 1.355 & 1.216 \\
\end{tabular}
\end{center}
\end{table}

The effective Wilson coefficients appear in the factorizable decay
amplitudes in the combinations $a_{2i}= {z}_{2i}^{\rm eff}+{1\over
N_c}{z}_{2i-1}^{\rm eff}$ and $a_{2i-1}= {z}_{2i-1}^{\rm
eff}+{1\over N_c}{z}^{\rm eff}_{2i}$ $(i=1,\cdots,5)$. For
$K\to\pi\pi$ decays, nonfactorizable effects in hadronic matrix
elements can be absorbed into the parameters $a_i^{\rm eff}$
\cite{Cheng94,Kamal94,Soares}:
\be
a_{2i}^{\rm eff}= {z}_{2i}^{\rm eff}+\left({1\over
N_c}+\chi_{2i}\right){z}_{2i-1}^{\rm eff}, \qquad\quad
a_{2i-1}^{\rm eff}= {z}_{2i-1}^{\rm eff}+\left({1\over
N_c}+\chi_{2i-1}\right){z}^{\rm eff}_{2i}, \label{aeff}
\en
with $\chi_i$ being the nonfactorizable terms. Likewise,
\be
b_{2i}^{\rm eff}= {y}_{2i}^{\rm eff}+\left({1\over
N_c}+\chi_{2i}\right){y}_{2i-1}^{\rm eff}, \qquad\quad
b_{2i-1}^{\rm eff}= {y}_{2i-1}^{\rm eff}+\left({1\over
N_c}+\chi_{2i-1}\right){y}^{\rm eff}_{2i}. \label{beff}
\en
To proceed, we shall assume that nonfactorizable effects in the
matrix elements of $(V-A)(V+A)$ operators differ from that of
$(V-A)(V-A)$ operators; that is,
\be
\label{chiLR} && \chi_{LL}\equiv \chi_1=\chi_2= \chi_3=
\chi_4=\chi_9=\chi_{10}, \non \\ && \chi_{LR}\equiv \chi_5=\chi_6=
\chi_7= \chi_8,
\en
and $\chi_{LR}\neq\chi_{LL}$. Theoretically, a primary reason is
that the Fierz transformation of the $(V-A)(V+A)$ operators
$O_{5,6,7,8}$ is quite different from that of $(V-A)(V-A)$
operators $O_{1,2,3,4}$ and $O_{9,10}$ \cite{CCTY}.
Experimentally, we have learned from nonleptonic charmless $B$
decays that $\chi_{LR}(B)\neq \chi_{LL}(B)$ \cite{CCTY,CY99}.  As
shown in \cite{Cheng99}, the nonfactorized term $\chi_{LL}$ can be
extracted from $K^+\to\pi^+\pi^0$ decay to be
\be
\label{chiLL} \chi_{LL}=-0.73\,.
\en
Contrary to the nonfactorized effects in hadronic charmless $B$
decays, which are dominated by hard gluon exchange in the heavy
quark limit \cite{Beneke} and expected to be small due to the
large energy released in the decay process, the nonfactorized term
in $K\to\pi\pi$ is large and nonperturbative in nature, namely it
comes mainly from soft gluon exchange.

\section{Non-perturbative parameters $B_{\lowercase{i}}$}
In the literature it is often to parametrize the hadronic matrix
elements in terms of the non-perturbative bag parameters
$B_i^{(0)}$ and $B_i^{(2)}$ which describe the scale and scheme
dependence of the hadronic matrix elements $\la Q_i(\mu)\ra$:
\be
B_i^{(0)}(\mu)\equiv {\la Q_i(\mu)\ra_0\over \la Q_i\ra_0^{\rm
VIA}},\qquad B_i^{(2)}(\mu)\equiv {\la Q_i(\mu)\ra_2\over \la
Q_i\ra_2^{\rm VIA}},
\en
where $\la Q_i\ra^{\rm VIA}$ denote the matrix elements evaluated
under the vacuum insertion approximation. In order to evaluate the
parameters $B_i^{(0,2)}$, as an example we consider the
vertex-type and penguin-type corrections to the hadronic matrix
element of the four-quark operator $Q_1$ in the NDR scheme
\cite{CCTY}:
\be
\label{Q1} \la Q_1(\mu)\ra &=& \left[1+{\alpha_s\over
4\pi}\left(-2\ln {\mu_f\over\mu}+3\right)\right]\la
Q_1(\mu_f)\ra+{\alpha_s\over 4\pi}\left(6\ln{\mu_f\over
\mu}-9\right)\la Q_2(\mu_f)\ra  \\ && -{\alpha_s\over
8\pi}\left(G(m_u,k,\mu)-{2\over 3}\right)\la
P(\mu_f)\ra-{\alpha\over 9\pi}\left(G(m_u,k,\mu)-{2\over
3}\right)\la Q_7(\mu_f)+Q_9(\mu_f)\ra, \non
\en
where
\be
P=-{1\over 3}Q_3+Q_4-{1\over 3}Q_5+Q_6.
\en
The parameters $B_1^{(0,2)}$ are then determined from Eq.
(\ref{Q1}). In general, the hadronic parameters
$B_i^{(0,2)}(\mu,\mu_f)$ have the expressions:\footnote{Note that
our convention for $Q_1,Q_2$ (and hence $B_1,B_2$) differs from
that in \cite{Buras93,Buras96} where the labels 1 and 2 are
interchanged.} \be \label{Bi} B_1^{(0)} &=& \Bigg\{ \la
Q_1\ra_{0}+{\alpha_s\over 4\pi}\left(\gamma_V\ln{\mu_f\over
\mu}+\hat r_V\right)_{1i}\la Q_i\ra_{0}  -{\alpha_s\over
8\pi}\left(G(m_u)-{2\over 3}\kappa\right)\la P\ra_{0}  \non \\ &&
-{\alpha\over 9\pi}\left(G(m_u)-{2\over 3}\kappa\right) \la
Q_7+Q_9\ra_{0}\Bigg\}\Big/\la Q_1\ra_{0}^{\rm VIA}, \non\\
B_1^{(2)} &=& \Bigg\{ \la Q_1\ra_{2}+{\alpha_s\over
4\pi}\left(\gamma_V\ln{\mu_f\over \mu}+\hat r_V\right)_{1i}\la
Q_i\ra_{2}\non \\ &&  -{\alpha\over 9\pi}\left(G(m_u)-{2\over
3}\kappa\right) \la Q_7+Q_9\ra_{2}\Bigg\}\Big/\la Q_1\ra_{2}^{\rm
VIA}, \non\\ B_2^{(0,2)} &=& \Bigg\{ \la
Q_2\ra_{0,2}+{\alpha_s\over 4\pi}\left(\gamma_V\ln{\mu_f\over
\mu}+\hat r_V\right)_{2i}\la Q_i\ra_{0,2} \non \\ && -{\alpha\over
3\pi} \left(G(m_u)-{2\over 3}\kappa\right)\la
Q_7+Q_9\ra_{0,2}\Bigg\}\Big/\la Q_2\ra_{0,2}^{\rm VIA}, \non \\
B_3^{(0)} &=& \Bigg\{ \la Q_3\ra_{0}+{\alpha_s\over
4\pi}\left(\gamma_V\ln{\mu_f\over \mu}+\hat r_V\right)_{3i}\la
Q_i\ra_{0} \non \\  && -{\alpha_s\over
4\pi}\left(G(m_d)+G(m_s)-{4\over 3}\kappa\right)\la
P\ra_{0}\Bigg\}\Big/\la Q_3\ra_{0}^{\rm VIA}, \non \\ B_4^{(0)}
&=& \Bigg\{ \la Q_4\ra_{0}+{\alpha_s\over
4\pi}\left(\gamma_V\ln{\mu_f\over \mu}+\hat r_V\right)_{4i}\la
Q_i\ra_{0}\non \\  && -{\alpha_s\over
4\pi}\left[G(m_u)+G(m_d)+G(m_s)\right]\la P\ra_{0}\Bigg\}\Big/\la
Q_4\ra_{0}^{\rm VIA}, \non \\ B_5^{(0)} &=& \Bigg\{ \la
Q_5\ra_{0}+{\alpha_s\over 4\pi}\left(\gamma_V\ln{\mu_f\over
\mu}+\hat r_V\right)_{5i}\la Q_i\ra_{0}\Bigg\}\Big/\la
Q_5\ra_{0}^{\rm VIA},   \non \\ B_6^{(0)} &=& \Bigg\{ \la
Q_6\ra_{0}+{\alpha_s\over 4\pi}\left(\gamma_V\ln{\mu_f\over
\mu}+\hat r_V\right)_{6i}\la Q_i\ra_{0}\non \\  && -{\alpha_s\over
4\pi}\left[G(m_u)+G(m_d)+G(m_s)\right]\la P\ra_{0}\Bigg\}\Big/\la
Q_6\ra_{0}^{\rm VIA}, \non \\ B_7^{(0,2)} &=& \Bigg\{ \la
Q_7\ra_{0,2}+{\alpha_s\over 4\pi}\left(\gamma_V\ln{\mu_f\over
\mu}+\hat r_V\right)_{7i}\la Q_i\ra_{0,2}\Bigg\}\Big/\la
Q_7\ra_{0,2}^{\rm VIA},   \non \\ B_8^{(0,2)} &=& \Bigg\{ \la
Q_8\ra_{0,2}+{\alpha_s\over 4\pi}\left(\gamma_V\ln{\mu_f\over
\mu}+\hat r_V\right)_{8i}\la Q_i\ra_{0,2}\Bigg\}\Big/\la
Q_8\ra_{0,2}^{\rm VIA},   \non \\ B_9^{(0,2)} &=& \Bigg\{ \la
Q_9\ra_{0,2}+{\alpha_s\over 4\pi}\left(\gamma_V\ln{\mu_f\over
\mu}+\hat r_V\right)_{9i}\la Q_i\ra_{0,2}\Bigg\}\Big/\la
Q_9\ra_{0,2}^{\rm VIA},   \non \\ B_{10}^{(0,2)} &=& \Bigg\{ \la
Q_{10}\ra_{0,2}+{\alpha_s\over 4\pi}\left(\gamma_V\ln{\mu_f\over
\mu}+\hat r_V\right)_{10i}\la Q_i\ra_{0,2}\Bigg\}\Big/\la
Q_{10}\ra_{0,2}^{\rm VIA}.
\en
For simplicity we have dropped the parameters $k$ and $\mu$ in the
argument of the function $G$. Note that the effective Wilson
coefficient $\tilde z_i^{\rm eff}$ is in general not equal to
$z_i(\mu)B_i(\mu)$, but the physical amplitude in terms of $\tilde
z_i^{\rm eff}$ or $z_i(\mu)B_i(\mu)$ is the same.

The $K\to\pi\pi$ matrix elements under the vacuum insertion
approximation read (see e.g., \cite{Buras93})
\be
\la Q_1\ra_0 &=& {1\over 3}X\left(2-{1\over N_c}\right), \qquad
\qquad \quad \la Q_1\ra_2 = {\sqrt{2}\over 3}X\left(1+{1\over
N_c}\right), \non \\ \la Q_2\ra_0 &=& {1\over 3}X\left(-1+{2\over
N_c}\right), \qquad \qquad~ \la Q_2\ra_2 = {\sqrt{2}\over
3}X\left(1+{1\over N_c}\right), \non \\ \la Q_3\ra_0 &=& {1\over
N_c}X, \qquad\qquad \qquad\qquad\quad \la Q_4\ra_0=X, \non \\ \la
Q_5\ra_0 &=& -{4\over N_c}\,\sqrt{3\over 2}\,v^2(f_K-f_\pi),
\qquad~ \la Q_6\ra_0 = -4\,\sqrt{3\over 2}\,v^2(f_K-f_\pi), \non\\
\la Q_7\ra_0 &=& {\sqrt{6}\over N_c}f_K v^2+{1\over 2}X, \qquad
\qquad \quad \la Q_7\ra_2 = {\sqrt{3}\over N_c}f_\pi v^2-{1\over
\sqrt{2}}X, \non\\ \la Q_8\ra_0 &=& \sqrt{6}f_K v^2+{1\over
2N_c}X, \qquad \qquad \la Q_8\ra_2 = \sqrt{3}f_\pi v^2-{1\over
N_c\sqrt{2}}X, \non\\ \la Q_9\ra_0 &=& -{1\over 2}X\left(1-{1\over
N_c}\right), \qquad\qquad~ \la Q_9\ra_2 = -{1\over
\sqrt{2}}X\left(1+{1\over N_c}\right), \non \\ \la Q_{10}\ra_0 &=&
{1\over 2}X\left(1-{1\over N_c}\right), \qquad \qquad\quad \la
O_{10}\ra_2 = {1\over \sqrt{2}}X\left(1+{1\over N_c}\right),
\label{O02me}
\en
where $X=\sqrt{3/2}\,f_\pi(m_K^2-m_\pi^2)$, and
\be
v={m^2_{\pi^\pm}\over m_u+m_d}={m^2_{K^0}\over
m_d+m_s}={m^2_K-m^2_\pi\over m_s-m_u}
\en
characterizes the quark-order parameter $\la \bar qq\ra$ which
breaks chiral symmetry spontaneously.

To evaluate $B_i^{(0,2)}$ we need to take into account
nonfactorized effects on hadronic matrix elements. As discussed in
Sec. II, this amounts to replacing $1/N_c$ by $1/N_c+\chi_{LL}$
for $(V-A)(V-A)$ quark operators and by $1/N_c+\chi_{LR}$ for
$(V-A)(V+A)$ operators. For example, \be \label{example} {\la
Q_1\ra_0\over \la Q_1\ra_0^{\rm VIA}} = 1-{3\over 5}\chi_{LL},
\qquad {\la Q_2\ra_0\over \la Q_2\ra_0^{\rm VIA}} = 1-6\chi_{LL},
\qquad {\la Q_5\ra_0\over \la Q_5\ra_0^{\rm VIA}} = 1+3\chi_{LR}.
\en
Although the nonfactorized term $\chi_{LL}$ is fixed by the
measurement of $K^+\to\pi^+\pi^0$ to be $-0.73$ \cite{Cheng99}, no
constraint on $\chi_{LR}$ can be extracted from $K^0\to\pi\pi$.
Nevertheless, we learned from hadronic charmless $B$ decays that
$\chi_{LR}\neq\chi_{LL}$ \cite{CCTY}. As shown in Fig. 1, the
parameters $B_5,~B_7^{(0)}$ and $B_7^{(2)}$ are quite sensitive to
$\chi_{LR}$, whereas $B_6,~B_7^{(0)}$ and $B_7^{(2)}$ stay stable.
Lattice calculations suggest that $B_5\simeq B_6=1.0\pm 0.2$ and
$B_7^{(2)}=0.6\pm 0.1$ at $\mu=2$ GeV \cite{Ciuchini}. The lattice
results roughly imply the constraint $-0.45<\chi_{LR}<0$. We shall
see in Sec. IV that the $A_0/A_2$ ratio and
$\varepsilon'/\varepsilon$ are not very sensitive to the variation
of $\chi_{LR}$. Using $m_u=3.5$ MeV, $m_d=7.0$ MeV, $m_s=140$ MeV
at $\mu=1$ GeV and $\chi_{LR}=-0.1$, the numerical values of
$B_i^{(0,2)}$ are listed in Tables II and III.

\begin{figure}[th]
\psfig{figure=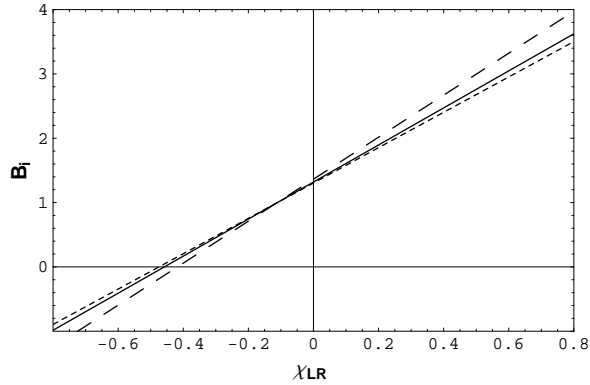,height=2.0in} \vspace{0.4cm}
    \caption{{\small The bag parameters $B_5$ (solid line),
    $B_7^{(0)}$ (dotted line) and $B_7^{(2)}$ (dashed line)
    evaluated in the NDR scheme
    versus $\chi_{LR}$ for $m_s(1\,{\rm GeV})=140$ MeV.}}
\end{figure}

\vskip 0.5cm
\begin{table}[ht]
\caption{Numerical values of the non-perturbative bag parameters
$B_i^{(0)}$ at $\mu=1$ GeV in NDR and HV schemes for $m_s(1\,{\rm
GeV}) =140$ MeV, $\mu_f=1$ GeV and $k^2=m_K^2/2$,
$\chi_{LL}=-0.73$ and $\chi_{ LR}=-0.1$. The results for
$B_i^{(0)}$ in the absence of nonfactorized contributions (i.e.
$\chi_{LL}=\chi_{LR}=0$) are shown in parentheses.}
\begin{center}
\begin{tabular}{ l c c c c c c c c c c c }
 & $B_1^{(0)}$ & $B_2^{(0)}$ & $B_3^{(0)}$ & $B_4^{(0)}$
 & $B_5^{(0)}$ & $B_6^{(0)}$ & $B_7^{(0)}$ & $B_8^{(0)}$
 & $B_9^{(0)}$ & $B_{10}^{(0)}$
 \\ \hline
 NDR & 2.5(1.8) & 8.7(3.1) & -0.1(2.8) & 2.7(2.7) & 1.0(1.3) & 1.5(1.5)
 & 1.0(1.3) & 1.6(1.6) & 3.1(1.5) & 3.1(1.5)  \\
 HV & 2.6(2.1) & 8.0(2.7) & 0.9(3.8) & 2.6(2.7) & 1.7(2.0) & 1.5(1.5)
 & 1.7(1.9) & 1.6(1.6) & 2.9(1.4) & 2.9(1.4) \\
\end{tabular}
\end{center}
\end{table}

\begin{table}[ht]
\caption{Same as Table II except for $B_i^{(2)}$.}
\begin{center}
\begin{tabular}{ l c c c c c c c }
 & $B_1^{(2)}$ & $B_2^{(2)}$ & $B_7^{(2)}$ & $B_8^{(2)}$
 & $B_9^{(2)}$ & $B_{10}^{(2)}$
 \\ \hline
 NDR & 0.34(0.75) & 0.34(0.74) & 1.0(1.4) & 1.6(1.6) & 0.35(0.76) & 0.35(0.76) \\
 HV & 0.37(0.81) & 0.36(0.79) & 1.8(2.1) & 1.6(1.6) & 0.37(0.81) & 0.37(0.81) \\
\end{tabular}
\end{center}
\end{table}

\section{$K\to\pi\pi$ isospin amplitudes and
$\varepsilon'/\varepsilon$} In terms of the effective Wilson
coefficients defined in Sec. II, the CP-even $\Delta I=1/2$ and
$\Delta I=3/2$ $K\to\pi\pi$ amplitudes have the form
\cite{Cheng99}:
\be
\label{A02} {\rm Re}A_0 & =& {G_F\over\sqrt{2}}\,{{\rm Re}
(V_{ud}V^*_{us})\over\cos\delta_0}\Bigg\{ \Big[\,{2\over
3}a_1-{1\over 3}a_2+a_4+{1\over 2}(a_7-a_9+a_{10})\Big]X  \non \\
&& \qquad \qquad\qquad\quad-
2\sqrt{6}\,v^2(f_K-f_\pi)a_6+\sqrt{6}\,v^2f_K a_8\Bigg\}, \non \\
{\rm Re}A_2 &=& {G_F\over\sqrt{2}}\,{ {\rm Re}(
V_{ud}V^*_{us})\over\cos\delta_2}\,{1\over 1-\Omega_{\rm IB}}
\Bigg\{ \Big[a_1+a_2+{3\over
2}(-a_7+a_9+a_{10})\Big]\,{\sqrt{2}\over 3} X+\sqrt{3}\,f_\pi
v^2a_8\Bigg\},
\en
where $\Omega_{\rm IB}\equiv A_2^{\rm IB}/A_2$ describes the
isospin breaking contribution to $K^+\to\pi^+\pi^0$ due to the
$\pi-\eta-\eta'$ mixing, and $\delta_0$ as well as $\delta_2$ are
S-wave $\pi\pi$ scattering isospin phase shifts. For simplicity,
we have dropped the superscript ``eff" of $a_i$ in Eq.
(\ref{A02}).

The direct CP-violation parameter $\varepsilon'/\varepsilon$ given
by the general expression
\be  \label{ep1}
{\varepsilon'\over
\varepsilon}=\,{\omega\over\sqrt{2}|\varepsilon|}\left( {{\rm
Im}A_2\over {\rm Re}A_2}-{{\rm Im}A_0\over {\rm Re}A_0}\right)
\en
can be recast in the form
\be
{\varepsilon'\over \varepsilon}={G_F\omega\over 2|\varepsilon|{\rm
Re}A_0}{\rm Im}(V_{td}V^*_{ts})&& \Bigg\{  {1
\over\cos\delta_0}\Big[(b_4+{1\over 2}b_7-{1\over 2}b_9+{1\over
2}b_{10})X  \non \\ && -2\sqrt{6}\,v^2(f_K-f_\pi)b_6
+\sqrt{6}\,v^2f_Kb_8\Big](1-\Omega_{\rm IB}) \non \\ &&
-{1\over\omega}{1\over\cos\delta_2}
\Big[(-b_7+b_9+b_{10})X/\sqrt{2}+\sqrt{3}\,v^2 f_\pi
b_8\Big]\Bigg\},
\en
where $\omega\equiv A_2/A_0=1/22.2$\,.

Alternatively, the $K\to\pi\pi$ amplitudes and direct CP violation
can be expressed in terms of the non-perturbative parameters
$B_i^{(0,2)}$:
\be
{\rm Re}A_0 &=& {G_F\over\sqrt{2}}\,{{\rm Re}(
V_{ud}V^*_{us})\over\cos\delta_0}\sum_{i=1}^{10} z_iB_i^{(0)}\la
Q_i\ra_0^{\rm VIA}, \non \\ {\rm Re}A_2 &=&
{G_F\over\sqrt{2}}\,{{\rm Re}(
V_{ud}V^*_{us})\over\cos\delta_2}\,{1\over 1-\Omega_{\rm IB}}
\sum_{i=1}^{10} z_iB_i^{(2)}\la Q_i\ra_2^{\rm VIA},
\en
and
\be \label{ep2}
{\varepsilon'\over
\varepsilon}={G_F\omega\over 2|\varepsilon|{\rm Re}A_0}{\rm
Im}(V_{td}V^*_{ts})\,\Bigg\{ && {1\over\cos\delta_0}\sum
_{i=3}^{10}y_iB_i^{(0)}\la Q_i\ra_0^{\rm VIA}(1-\Omega_{\rm IB})
\non \\ && -{1\over\omega}{1\over\cos\delta_2}\sum
_{i=3}^{10}y_iB_i^{(2)}\la Q_i\ra_2^{\rm VIA}\Bigg\}.
\en
We have checked explicitly that the numerical values of $\Delta
I=1/2,~3/2$ amplitudes and $\varepsilon'/\varepsilon$ obtained
using the parameters $B_i^{(0,2)} $ given in Tables II and III are
the same as that calculated using the effective Wilson
coefficients $\tilde z_i^{\rm eff}$ and $\tilde y_i^{\rm eff}$
shown in Table I, as it should be. Since $z_i^{\rm eff}$ and
$y_i^{\rm eff}$ show a better scale independence than $\tilde
z_i^{\rm eff}$ and $\tilde y_i^{\rm eff}$, we will use the former
set of effective Wilson coefficients in ensuing
calculations.\footnote{The results of $A_0/A_2$ and
$\varepsilon'/\varepsilon$ calculated using the bag parameters
$B_i^{(0,2)} $ and effective Wilson coefficients $z_i^{\rm eff}$
and $y_i^{\rm eff}$ are numerically very similar except for
$\varepsilon'/\varepsilon$ in the HV scheme which is slightly
small in terms of $B$-parameters by around 15\% compared to that
evaluated in terms of $y_i^{\rm eff}$. For example, Eq.
(\ref{ep2}) leads to $\varepsilon'/\varepsilon=0.59\times 10^{-3}$
at $m_s({\rm 1\,GeV})=150$ MeV in the HV scheme, while Eq.
(\ref{A02}) yields $\varepsilon'/\varepsilon=0.70\times 10^{-3}$
[see Eq. (\ref{epvalue})] in the same scheme.}

Using $\Omega_{\rm IB}=0.25\pm 0.02$ \cite{Cheng99},
$\delta_0=(34.2\pm 2.2)^\circ,~ \delta_2=-(6.9\pm 0.2)^\circ$
\cite{Chell}, $\chi_{LL}=-0.73$ and $\chi_{LR}=-0.1$, we plot in
Fig. 2 the ratio $A_0/A_2$ as a function of $m_s$ at the
renormalization scale $\mu=1$ GeV. Specifically, we obtain
\be
\label{A0/A2} {{\rm Re}A_0\over {\rm Re}A_2}=\cases{ 15.2 &
at~$m_s\,(1\,{\rm GeV})=125$ MeV, \cr 13.6 & at~$m_s\,(1\,{\rm
GeV})=150$ MeV, \cr 12.7 & at~$m_s\,(1\,{\rm GeV})=175$ MeV. }
\en
It is clear that the strange quark mass is favored to be smaller
and that the prediction is renormalization scheme independent, as
it should be. In Fig. 3 we study the dependence of $A_0/A_2$ on
the unknown nonfactorized term $\chi_{LR}$. It turns out that the
ratio decreases slowly with $\chi_{LR}$, but it stays stable
within the allowed region $-0.45<\chi_{LR}<0$.

\begin{figure}[th]
\psfig{figure=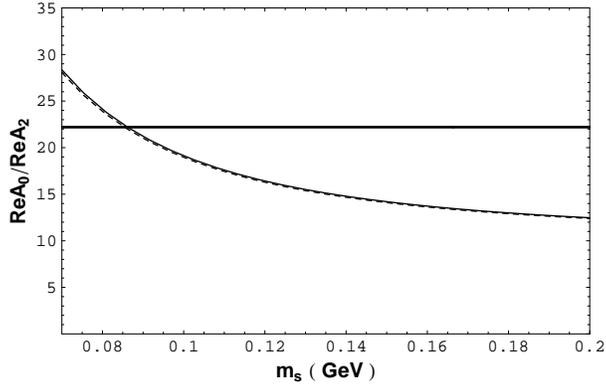,height=2.0in} \vspace{0.4cm}
    \caption{{\small The ratio of ${\rm Re}A_0/{\rm Re}A_2$ versus
    $m_s$ (in units of
    GeV) at the renormalization scale $\mu=1$ GeV,
    where the solid (dotted)
    curve is calculated in the NDR (HV) scheme and use of
    $\chi_{LR}=-0.1$ has been made. The solid thick line
    is the experimental value for Re$A_0/{\rm Re}A_2$.}}
\end{figure}

For direct CP violation, we find for
Im$(V_{td}V^*_{ts})=1.29\times 10^{-4}$ (see Fig. 4) \be
\label{epvalue} {\varepsilon'\over\varepsilon}=\cases{
1.56~(1.02)\times 10^{-3} & at~$m_s\,(1\,{\rm GeV})=125$ MeV, \cr
1.07~(0.70)\times 10^{-3} & at~$m_s\,(1\,{\rm GeV})=150$ MeV, \cr
0.78~(0.51)\times 10^{-3} & at~$m_s\,(1\,{\rm GeV})=175$ MeV, }
\en
in the NDR scheme, where the calculations in the HV scheme are
shown in parentheses. Experimentally, the world average including
NA31 \cite{NA31}, E731 \cite{E731}, KTeV \cite{KTeV} and NA48
\cite{NA48} results is
\be
{\rm Re}(\varepsilon'/\varepsilon)=(2.13\pm 0.46)\times 10^{-3}.
\en

\begin{figure}[t]
\psfig{figure=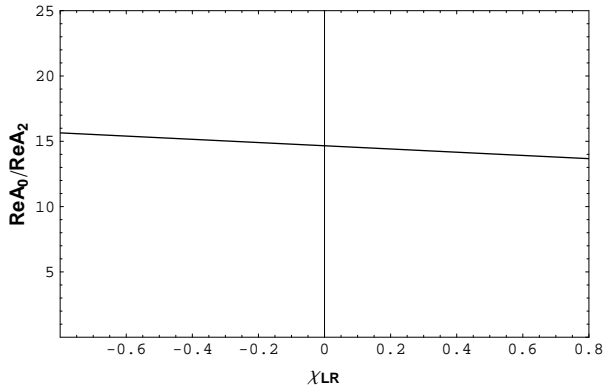,height=2.0in} \vspace{0.4cm}
    \caption{{\small The ratio of ${\rm Re}A_0/{\rm Re}A_2$
    versus $\chi_{LR}$ for
    $m_s(1\,{\rm GeV})=140$ MeV.}}
\end{figure}

\begin{figure}[th]
\psfig{figure=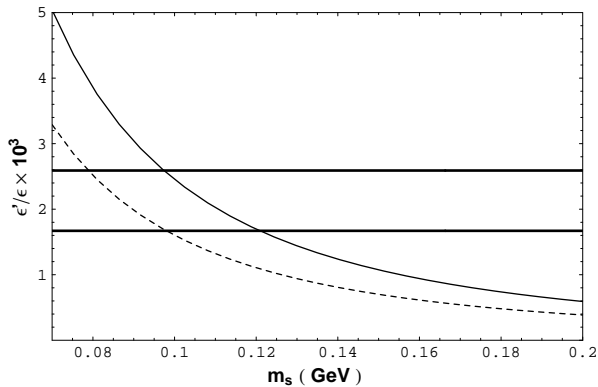,height=2.0in} \vspace{0.4cm}
    \caption{{\small Direct CP violation $\varepsilon'/\varepsilon$
    versus $m_s$ (in units of GeV) at the renormalization scale $\mu=1$ GeV,
    where the solid (dotted)
    curve is calculated in the NDR (HV) scheme and use of Im$(V_{td}V^*_{ts})
    =1.29\times 10^{-4}$ and
    $\chi_{LR}=-0.1$ has been made. The solid thick
    lines are the world average value for $\varepsilon'/\varepsilon$ with
    one sigma errors.}}
\end{figure}

\section{Discussions}
\subsection{Bag parameters $B_i$}
In Sec. III we have computed the non-perturbative parameters
$B_i^{(0,2)}$ at $\mu=1$ GeV in NDR and HV schemes. Our results
$B_{1,2}^{(2)}=0.34$ and $B_{9,10}^{(2)}=0.35$ in the NDR scheme
are smaller than the value 0.48 quoted in \cite{Buras93} for
$\mu=1.3$ GeV. This is because we have taken into account isospin
breaking contributions to the $\Delta I=3/2$ amplitude so that
Re$A_2$ is enhanced by a factor of $1/(1-\Omega_{\rm IB})$ [see
Eq. (\ref{A02})]. Consequently, it is necessary to impose large
nonfactorized effects and hence small $B_{1,2}^{(2)}$ to suppress
$A_2$. We observe from Tables II and III that the parameter
$B_2^{(0)}$ has the largest deviation from unity; it is equal to
8.7 in the NDR scheme. This is mainly because the ratio $\la
Q_2\ra_0/\la Q_2\ra_0^{\rm VIA}=1-6\chi_{LL}$ [cf. Eq.
(\ref{example})] is greatly enhanced by the nonfactorized effect,
recalling that $\chi_{LL}=-0.73\,$. Note that our results for
$B_{1,2}^{(0)}$ are close to that obtained in the chiral quark
model \cite{Antonelli}.

There exist some nonperturbative calculations for $B_6^{(0)}$ and
$B_{7,8}^{(2)}$. Among them, lattice calculations are carried out
at $\mu=2$ GeV and in the NDR scheme. However, the lattice results
for $B_{7,8}^{(2)}$ are much more reliable and solid than
$B_6^{(0)}$. Most approaches find $B_8^{(2)}$ below unity and
$B_8^{(2)}< B_6^{(0)}$, while we obtain $B_8^{(2)}=1.6$ and
$B_8^{(2)}\sim B_6^{(0)}$. Since the radiative correction
$(\alpha_s/4\pi)[(\hat r)_{87}\la Q_7\ra_2+(\hat r)_{88}\la
Q_8\ra_2]/\la Q_8\ra_2^{\rm VIA}$ is positive in our case [see Eq.
(\ref{Bi})], it is not possible to push $B_8^{(2)}$ down below
unity.

As for the scheme dependence of $B_i$, it was argued in
\cite{Bosch} that $B_6^{(0)}({\rm HV})\approx 1.2\,B_6^{(0)}({\rm
NDR})$ and $B_8^{(2)}({\rm HV})\approx 1.2\,B_8^{(2)}({\rm NDR})$,
whereas we find a very weak scheme dependence for $B_6^{(0)}$ and
$B_8^{(2)}$ but strong dependence for $B_5^{(0)}$ and
$B_7^{(0,2)}$: $B_5^{(0)}({\rm HV})=1.7\,B_5^{(0)}({\rm NDR})$,
$B_7^{(0)}({\rm HV})=1.7\,B_7^{(0)}({\rm NDR})$ and
$B_7^{(2)}({\rm HV})=1.8\,B_7^{(2)}({\rm NDR})$. We have also
studied the $m_s$ dependence of $B$-parameters and found that only
$B_{1,3,4}^{(0)}$ exhibit a significant $m_s$ dependence, while
the other $B$-parameters are nearly $m_s$ independent. For
example, we obtain $B_1^{(0)}=3.2$ if $m_s({\rm 1\,GeV})=100$ MeV
and $B_1^{(0)}=2.5$ if $m_s({\rm 1\,GeV})=140$ MeV (see Table II).
Note that some other models predict a different $m_s$ behavior for
$B$-parameters. For example, $B_6^{(0)}$ is proportional to $m_s$
in the chiral quark model \cite{Antonelli}.

\subsection{$K\to\pi\pi$ amplitudes}
From Fig. 2 or Eq. (\ref{A0/A2}) we see that about (60-70)\% of
Re$A_0$ amplitude is accounted for in the present approach if
$m_s$(1\,GeV) lies in the range (125-175)\,MeV. Specifically,
$Q_1$, $Q_2$ and penguin operators explain 66\%, 18\% and 14\%,
respectively, of the $A_0$ amplitude for $m_s(1\,{\rm GeV})=150$
MeV. Hence, tree-level current$\times$current operators account
for around 85\% of Re$A_0$. However, contrary to \cite{Bijnens},
we find that penguin-like diagrams induced by $Q_1$, i.e., the
penguin operators in Eq. (\ref{Q1}), contributes only about 15\%
to Re$A_0$. As conjectured in \cite{Cheng99}, the $W$-exchange
mechanism could provide an additional important enhancement of the
$A_0$ amplitude. Since the $W$-exchange amplitude in charmed meson
decay is comparable to the internal $W$-emission one \cite{CC}, it
is conceivable that in kaon physics the long-distance contribution
to $W$-exchange is as important as the external $W$-emission
amplitude.

It is instructive to see how the predictions of Re$A_0$ and
Re$A_2$ amplitudes and the $\Delta I=1/2$ rule progress at various
stages. In the absence of QCD corrections, we have $a_2={1\over
3}a_1$ and $a_3=a_4\cdots=a_{10}=0$ under the vacuum insertion
approximation. It follows from Eq. (\ref{A02}) that \cite{Cheng89}
\be
{{\rm Re}A_0\over {\rm Re}A_2}=\,{5\over 4\sqrt{2}}=0.9 \quad{\rm
(in~ absence~of~QCD~corrections)}. \label{naive}
\en
With the inclusion of lowest-order short-distance QCD corrections
to the Wilson coefficients $z_1$ and $z_2$ evaluated at $\mu=1$
GeV, $A_0/A_2$ is enhanced from the value of 0.9 to 2.0, and it
becomes 2.3 if $m_s(1\,{\rm GeV})=150$ MeV and QCD penguin as well
as electroweak penguin effects are included. This ratio is
suppressed to 1.7 with the inclusion of the isospin-breaking
effect, but it is increased again to the value of 2.0 in the
presence of final-state interactions with $\delta_0=34.2^\circ$
and $\delta_2=-6.9^\circ$. At this point, we have Re$A_0=7.7\times
10^{-8}$\,GeV and Re$A_2=3.8\times 10^{-8}$\,GeV.  Comparing with
the experimental values
\be
{\rm Re}\,A_0=3.323\times 10^{-7}\,{\rm GeV}, \qquad {\rm Re}\,
A_2=1.497\times 10^{-8}\,{\rm GeV}, \label{expt}
\en
we see that the conventional calculation based on the effective
Hamiltonian and naive factorization predicts a too small $\Delta
I=1/2$ amplitude by a factor of 4.3 and a too large $\Delta I=3/2$
amplitude by a factor of 2.5\,. In short, it is a long way to go
to achieve the $\Delta I=1/2$ rule within the conventional
approach.

Replacing $c_i^{\rm LO}(\mu)$ by the effective Wilson coefficients
$c_i^{\rm eff}$, or equivalently replacing the LO Wilson
coefficients by the NLO ones and including vertex-like and
penguin-like corrections to four-quark operators, we obtain
Re$A_0=1.37\times 10^{-7}$\,GeV and Re$A_2=3.3\times 10^{-8}$\,GeV
and Re$A_0/{\rm Re}A_2=4.2$\,. Finally, the inclusion of
nonfactorized effects on hadronic matrix elements will enhance
Re$A_0/{\rm Re}A_2$ to the value of 13.8 with Re$A_0=2.07\times
10^{-8}$\,GeV and Re$A_2=1.50\times 10^{-8}$\,GeV. To summarize,
the enhancement of the ratio Re$A_0/{\rm Re}A_2$ is due to the
cumulative effects of the short-distance Wilson coefficients,
penguin operators, final-state interactions, nonfactorized effects
due to soft-gluon exchange, and radiative corrections to the
matrix elements of four-quark operators. Among them, the last two
effects, which are usually not addressed in previous studies (in
particular, the last one), play an essential role for explaining
the bulk of the $\Delta I=1/2$ rule.

\subsection{Direct CP violation $\varepsilon'/\varepsilon$}
From Fig. 4 or Eq. (\ref{epvalue}) we observe that, contrary to
the case of $A_0/A_2$, the prediction of
$\varepsilon'/\varepsilon$ shows some scale dependence (see Fig.
4); roughly speaking, $(\varepsilon'/\varepsilon)_{\rm NDR}\approx
1.5\,(\varepsilon'/\varepsilon)_{\rm HV}$. To understand this, we
note that $\varepsilon'/\varepsilon$ is dominated by $b_6$ and
$b_8$ terms (or $y_6^{\rm eff}$ and $y_8^{\rm eff}$) or
equivalently by the hadronic parameters $B_6^{(0)}$ and
$B_8^{(2)}$ (see Eqs. (\ref{ep1}), (\ref{ep2}) and Tables I-III).
Moreover, direct CP violation involves a large cancellation
between the dominant $y_6^{\rm eff}$ and $y_8^{\rm eff}$ terms.
The scale dependence of the predicted $\varepsilon'/\varepsilon$
is traced back to the scale dependence of the effective Wilson
coefficient $y_6^{\rm eff}$ (see Table I). As mentioned before,
formally $y_6^{\rm eff}$ should be scale independent to the order
$\alpha_s$. It is thus not clear to us why $y_6^{\rm eff}({\rm
NDR})$ and $y_6^{\rm eff}({\rm HV})$ are not the same to the
accuracy under consideration. Furthermore, the scale dependence of
$y_6^{\rm eff}$ is amplified by the strong cancellation between
QCD penguin and electroweak penguin contributions, which makes it
difficult to predict $\varepsilon'/\varepsilon$ accurately. It
appears to us that the different results of
$\varepsilon'/\varepsilon$ in  NDR and HV schemes can be regarded
as the range of theoretical uncertainties. It is easily seen that
a suppression of $B_8^{(2)}$ or an enhancement of $B_6^{(0)}$ will
render $\varepsilon'/\varepsilon$ larger; that is, a ratio of
$B_6^{(0)}/B_8^{(2)}$ greater than unity will help get a large
$\varepsilon'/\varepsilon$. However, in our approach
$B_8^{(2)}\sim B_6^{(0)}=1.5$ and they are nearly scheme
independent. We have also studied the dependence of
$\varepsilon'/\varepsilon$ on the nonfactorized effect $\chi_{LR}$
and found that it increases slowly with $\chi_{LR}$ (see Fig. 5),
opposite to the case of $A_0/A_2$.

\begin{figure}[th]
\psfig{figure=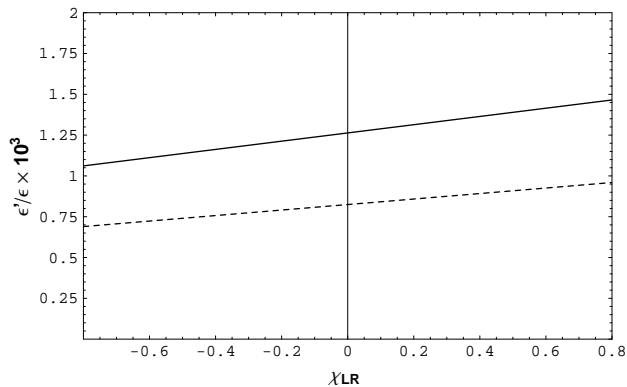,height=2.0in} \vspace{0.4cm}
    \caption{{\small Direct CP violation $\varepsilon'/\varepsilon$
   versus $\chi_{LR}$ for Im$(V_{td}V^*_{ts})=1.29\times 10^{-4}$
   and
    $m_s(1\,{\rm GeV})=140$ MeV, where the solid (dotted)
    curve is calculated in the NDR (HV) scheme.}}
\end{figure}

Since the $\Delta I=1/2$ rule and $\varepsilon'/\varepsilon$ are
both under-estimated theoretically, it is natural to ask if there
exists a strong correlation between them. The two principal
mechanisms responsible for the enhancement of $A_0/A_2$ are the
vertex-type as well as penguin-type corrections to the matrix
elements of four-quark operators, and the nonfactorized effect due
to soft-gluon exchange. Turning off these two effects by setting
$\chi_{LL}=\chi_{LR}=0$ and $y_i^{\rm eff}\to y_i^{\rm LO}$, we
find that $\varepsilon'/\varepsilon$ does not get changed in a
significant way. On the other hand, if a small strange quark mass
is responsible for the remaining enhancement necessary for
accommodating the data of $A_0/A_2$, it turns out that $m_s({\rm
1\,GeV})=85$ MeV and $\varepsilon'/\varepsilon=(2.3-3.5)\times
10^{-3}$. However, this $m_s$ is too small even compared to the
recent lattice result \cite{Ryan} which favors a lower strange
quark mass: $m_s({\rm 2\,GeV})=(84\pm 7)$\,MeV.

\section{Conclusions}
The $\Delta I=1/2$ rule and direct CP violation
$\varepsilon'/\varepsilon$ in kaon decays are studied within the
framework of the effective Hamiltonian approach in conjunction
with generalized factorization for hadronic matrix elements. Our
results are as follows.

\begin{enumerate}
\item We identify two principal sources responsible for the
enhancement of Re$A_0/{\rm Re}A_2$: the vertex-type as well as
penguin-type corrections to the matrix elements of four-quark
operators, which render the physical amplitude renormalization
scale and scheme independent, and nonfactorized effect due to
soft-gluon exchange, which is needed to suppress the $\Delta
I=3/2$ $K\to\pi\pi$ amplitude. This approach is not only much
simpler and logical than chiral loop calculations but also
applicable to heavy meson decays.

\item We obtain renormalization-scheme independent predictions for
$K\to\pi\pi$ amplitudes and find ${\rm Re}A_0/{\rm Re}A_2=13-15$
if $m_s$(1\,GeV) lies in the range (125$-$175)\,MeV. The
tree-level current$\times$current operators account for around
85\% of Re$A_0$. We conjecture that the $W$-exchange mechanism may
provide an additional important enhancement of the $\Delta I=1/2$
amplitude.

\item The bag parameters $B_i$, which are often employed to
parametrize the scale and scheme dependence of hadronic matrix
elements, are calculated in two different renormalization schemes
by considering the vertex-like and penguin-like corrections to
four-quark operators. It is found that $B_8^{(2)}\sim B_6^{(0)}$,
both of order 1.5 at $\mu=1$ GeV, are nearly $\gamma_5$ scheme
independent, whereas $B^{(0)}_{3,5,7}$ and $B_7^{(2)}$ show a
sizable scheme dependence. Our results $B_{1,2}^{(2)}=0.34$ and
$B_{9,10}^{(2)}=0.35$ in the NDR scheme are smaller than the value
quoted in the literature since we have taken into account isospin
breaking contributions to the $\Delta I=3/2$ amplitude. As for the
dependence of $B$-parameters on $m_s$, only $B_{1,3,4}^{(0)}$
exhibit a significant $m_s$ dependence, while the rest
$B$-parameters are almost $m_s$ independent.

\item Nonfactorizable contributions to the hadronic matrix elements
of $(V-A)(V-A)$ four-quark operators are extracted from the
measured $K^+\to\pi^+\pi^0$ decay to be $\chi_{LL}=-0.73$, while
the nonfactorized term for $(V-A)(V+A)$ operators lies in the
range $-0.45<\chi_{LR}<0$. We found that Re$A_0/{\rm Re}A_2$
($\varepsilon'/\varepsilon$)  decreases (increases) slowly with
$\chi_{LR}$.

\item For direct CP violation, the prediction of
$\varepsilon'/\varepsilon$ is renormalization scheme dependent
owing to the scale dependence with the effective Wilson
coefficient $y_6^{\rm eff}$ for reasons not clear to us. We obtain
$\varepsilon'/\varepsilon=(0.7-1.1)\times 10^{-3}$ if $m_s(1\,{\rm
GeV})=150$ MeV and $\varepsilon'/\varepsilon=(1.0-1.6)\times
10^{-3}$ if $m_s$ is as small as indicated by recent lattice
results.

\end{enumerate}

\vskip 0.5cm
\acknowledgements  This work is supported in part by
the National Science Council of the Republic of China under Grant
No. NSC89-2112-M001-016.

\end{document}